\documentclass[showkeys,superscriptaddress,floatfix,prd,10pt,aps,twocolumn]{revtex4-2}
\usepackage{graphicx,epstopdf}
\usepackage[caption=false]{subfig}
\usepackage{appendix}
\usepackage[T1]{fontenc}
\usepackage{lmodern}
\usepackage[dvipsnames,x11names]{xcolor}
\usepackage[colorlinks=true,linkcolor=NavyBlue,citecolor=ForestGreen,urlcolor=NavyBlue]{hyperref}
\usepackage[sort&compress]{natbib}
\usepackage{lipsum}
\usepackage{morefloats}
\usepackage[pdf]{pstricks}
\usepackage{amsmath}
\usepackage{amssymb}
\usepackage{amsfonts}
\usepackage{rotating}
\usepackage{cancel}
\usepackage{mathtools}
\usepackage{bbm}
\usepackage{dsfont}
\usepackage{bbold}
\usepackage{multirow}
\usepackage{ulem}
\usepackage{physics}
\usepackage{orcidlink}
\usepackage{colortbl}

\def\pslash{p\!\!\!\slash }
\def\qslash{q\!\!\!\slash }
\def\xslash{x\!\!\!\slash }
\def\eslash{\varepsilon\!\!\!\slash }

\begin{document}

\title{Hidden-charm pentaquarks: Electromagnetic structure in a diquark–diquark–antiquark model}

\author{Ula\c{s} \"{O}zdem\orcidlink{0000-0002-1907-2894}}%
\email[]{ulasozdem@aydin.edu.tr}
\affiliation{Health Services Vocational School of Higher Education, Istanbul Aydin University, Sefakoy-Kucukcekmece, 34295 Istanbul, T\"{u}rkiye}


 \begin{abstract}
  We systematically investigate the electromagnetic properties of exotic states whose internal structures remain uncertain and for which different models have been proposed. In this work, we focus on the magnetic dipole moments of hidden-charm pentaquark states using QCD light-cone sum rules with four distinct interpolating currents. The analysis accounts for contributions from both light and charm quark sectors, as well as higher-dimensional operators, ensuring convergence of the operator product expansion and dominance of the ground-state pole. Our results demonstrate a strong dependence of the magnetic moments on the internal quark configurations and spin alignments, revealing substantial variations among the different currents despite identical quark content and quantum numbers. Comparisons with existing studies indicate that while molecular-type predictions show general agreement, compact configurations yield markedly different values, including significant differences in sign and magnitude. These findings therefore underscore the sensitivity of electromagnetic observables to the internal structure of exotic hadrons and highlight their potential as probes to discriminate between competing structural models for spin–parity assignments and underlying quark dynamics.
  \end{abstract}


\maketitle

 \section{Introduction}\label{motivation}

The exploration of hadronic matter has, since the proposal of the quark model, extended well beyond conventional mesons and baryons to include more complex configurations such as tetraquarks, hybrids, glueballs, and pentaquarks. Quantum Chromodynamics (QCD) does not forbid the existence of such exotic systems, and their potential presence has remained a compelling question within hadron physics for decades. For a long period, however, these states were discussed largely on theoretical grounds until experimental confirmation finally arrived. In 2003, the Belle Collaboration reported the discovery of the $X(3872)$ resonance~\cite{Belle:2003nnu}, which was interpreted as a strong candidate for a tetraquark state. This milestone provided the first concrete evidence for multiquark dynamics and marked a turning point in the field. Since then, a growing list of exotic candidates has been observed by various collaborations, steadily broadening our understanding of the hadronic spectrum. A second decisive step came in 2015, when the LHCb Collaboration identified pentaquark-like structures in the $J/\psi p$ channel, revealing two states, $P_{\psi}^{N}(4380)$ and $P_{\psi}^{N}(4450)$~\cite{LHCb:2015yax}. This discovery provided the first strong evidence for hadrons containing five valence quarks. Subsequent studies with larger datasets refined the picture. The 2019 LHCb analysis demonstrated that the $P_{\psi}^{N}(4450)$ resonance actually corresponds to two nearby states, $P_{\psi}^{N}(4440)$ and $P_{\psi}^{N}(4457)$, and also uncovered a new peak associated with the $P_{\psi}^{N}(4312)$ state~\cite{LHCb:2019kea}. The status of the $P_{\psi}^{N}(4380)$, however, remains unresolved, as later analyses neither confirmed nor decisively excluded it.

Further discoveries continued to enrich the spectrum. In 2020, LHCb reported a new hidden-charm pentaquark, $P_{\psi s}^{\Lambda}(4459)$, in the $J/\psi \Lambda$ invariant mass spectrum~\cite{LHCb:2020jpq}. This was followed in 2022 by the observation of another state, $P_{\psi s}^{\Lambda}(4338)$, identified in $B^- \to J/\psi \Lambda p$ decays~\cite{LHCb:2022ogu}. More recently, the Belle Collaboration searched for $P_{\psi}^{N}$ states in $\Upsilon(1S,2S)$ inclusive decays to $pJ/\psi$ final states, but found no significant signal for $P_{\psi}^{N}(4312)$, $P_{\psi}^{N}(4440)$, or $P_{\psi}^{N}(4457)$~\cite{Belle:2024mcb}. Nevertheless, in 2025, Belle reported evidence for the $P_{\psi s}^{\Lambda}(4459)$ with a significance of $3.3\sigma$, including systematic and statistical uncertainties~\cite{Belle:2025pey}.  
A summary of the reported hidden-charm pentaquark states, including their measured properties and decay channels, is presented in Table~\ref{pentaquarks}. Alongside these observations, ongoing searches are probing states with higher strangeness. In particular, the CMS Collaboration recently studied the decay $\Lambda_b^0 \to J/\psi \Xi^- K^+$~\cite{CMS:2024vnm}. Although the data sample revealed no distinct resonance structure in the $J/\psi \Xi^-$ spectrum due to limited statistics and resolution, such analyses remain crucial for clarifying the mechanisms underlying beauty baryon decays and the production channels of multiquark states.
\begin{widetext}

\begin{table}[htp]
\caption{Reported hidden-charm pentaquark states by the LHCb \cite{LHCb:2015yax, LHCb:2019kea, LHCb:2020jpq, LHCb:2022ogu} and Belle \cite{Belle:2025pey} collaborations.}\label{pentaquarks}
\begin{tabular}{l|cccc}
\toprule 
\\
State  & Mass (MeV) & Width (MeV) & Quark content &  Observed decays \\
\\
\toprule
$P_{\psi}^{N}(4312)^+$ \cite{LHCb:2019kea}       & ~~$4311.9\pm0.7^{~+6.8}_{~-0.6}$~~  &  $9.8\pm2.7^{~+3.7}_{~-4.5}$    & $uudc\bar{c}$ & $\Lambda_b^0 \to J/\psi pK^-$ \\
$P_{\psi}^{N}(4440)^+$ \cite{LHCb:2019kea}      &    $4440.3\pm1.3^{~+4.1}_{~-4.7}$   &  $20.6\pm4.9^{~+8.7}_{~-10.1}$  & $uudc\bar{c}$& $\Lambda_b^0 \to J/\psi pK^-$ \\
$P_{\psi}^{N}(4457)^+$  \cite{LHCb:2019kea}    &    $4457.3\pm0.6^{~+4.1}_{~-1.7}$   &  $6.4\pm2.0^{~+5.7}_{~-1.9}$    & $uudc\bar{c}$ & $\Lambda_b^0 \to J/\psi pK^-$ \\
$P_{\psi s}^{\Lambda}(4459)^0$ \cite{LHCb:2020jpq}   &    $4458.8\pm2.9^{~+4.7}_{~-1.1}$   &  $17.3\pm6.5^{~+8.0}_{~-5.7}$   & $udsc\bar{c}$ &~$\Xi_b^- \to J/\psi \Lambda K^-$ \\
$P_{\psi s}^{\Lambda}(4338)^0$ \cite{LHCb:2022ogu}  &    $4338.2 \pm 0.7 \pm 0.4$   &  $7.0 \pm 1.2 \pm 1.3$    & $udsc\bar{c}$ & $B^-\to J/\psi \Lambda \bar{p}$\\
$P_{\psi s}^{\Lambda}(4459)^0$ \cite{Belle:2025pey}   &    $4471.7 \pm 4.8 \pm 0.6$   &  $21.9 \pm 13.1 \pm 2.7$   & $udsc\bar{c}$ &~$\Upsilon (1S, 2S)\to J/\psi \Lambda $ \\
\toprule
\end{tabular}
\end{table}

\end{widetext}

A variety of theoretical scenarios have been advanced to explain the nature of the observed pentaquark states. These include descriptions as compact, tightly bound five-quark configurations, loosely bound meson–baryon molecular states, and kinematic effects such as re-scattering processes. Comprehensive discussions of these possibilities can be found in Refs.~\cite{Esposito:2014rxa, Esposito:2016noz, Olsen:2017bmm, Lebed:2016hpi, Nielsen:2009uh, Brambilla:2019esw, Agaev:2020zad, Chen:2016qju, Ali:2017jda, Guo:2017jvc, Liu:2019zoy, Yang:2020atz, Dong:2021juy, Dong:2021bvy, Chen:2022asf, Meng:2022ozq, Liu:2024uxn, Wang:2025sic}. 
Despite extensive experimental and theoretical progress since their initial discovery, key questions regarding the internal structure of these states remain unresolved. In particular, their underlying composition, dynamical origin, and fundamental parameters such as spin–parity assignments are still subjects of active investigation and demand further clarification. 
Beyond the established states $P_{\psi}^{N}(4312)$, $P_{\psi}^{N}(4440)$, $P_{\psi}^{N}(4457)$, $P_{\psi s}^{\Lambda}(4338)$, and $P_{\psi s}^{\Lambda}(4459)$, ongoing searches aim to uncover additional candidates. Special attention is devoted to the possible existence of pentaquarks containing one or two strange quarks, as well as non-strange counterparts. Identifying and characterizing such states would provide critical insights into the spectrum of exotic hadrons and the mechanisms governing their formation.

Electromagnetic observables—such as magnetic dipole, electric quadrupole, and higher-order multipole moments—constitute powerful probes of hadronic substructure. These quantities are directly linked to the spatial distribution of quarks and to the alignment of their spins inside the hadron, thereby encoding essential information about its internal dynamics. In the case of hidden-charm pentaquarks, electromagnetic multipole moments represent a particularly sensitive diagnostic tool: they can illuminate the underlying quark–gluon composition, clarify spin–parity quantum numbers, and even provide hints about the geometric configuration of these exotic systems. 
Motivated by this, the present work is devoted to a QCD light-cone sum rule analysis~\cite{Chernyak:1990ag, Braun:1988qv, Balitsky:1989ry} of the magnetic dipole moments of hidden-charm pentaquarks. We construct four independent interpolating currents of the diquark–diquark–antiquark type, consistent with quantum numbers $\mathrm{J^P = \tfrac{1}{2}^-}$, and employ them to extract predictions for the corresponding magnetic dipole moments. These predictions aim to shed light on the structural features of the pentaquarks and serve as benchmarks for forthcoming experimental tests. In this way, our analysis contributes both to clarifying the role of multiquark dynamics in QCD and to enriching the phenomenology of exotic hadrons. 
It is important to note that, despite their potential significance, the electromagnetic characteristics of hidden-charm pentaquarks remain only sparsely explored. Theoretical investigations addressing multipole moments in hidden-charm or hidden-bottom systems are relatively limited, with existing studies including Refs.~\cite{Ozdem:2025jda, Wang:2016dzu, Ozdem:2018qeh, Ortiz-Pacheco:2018ccl, Xu:2020flp, Ozdem:2021btf, Ozdem:2021ugy, Li:2021ryu, Ozdem:2023htj, Wang:2023iox, Ozdem:2022kei, Gao:2021hmv, Ozdem:2024rch, Guo:2023fih, Ozdem:2022iqk, Wang:2022nqs, Wang:2022tib, Ozdem:2024jty, Li:2024wxr, Li:2024jlq, Ozdem:2024yel, Ozdem:2024rqx, Mutuk:2024ltc, Mutuk:2024jxf, Ozdem:2024usw, Ozdem:2025fks, Zhu:2025abk, Ozdem:2025ncd}. Our study therefore addresses an important gap in the current literature.

The remainder of this paper is structured as follows. In Section~\ref{formalism}, we outline the theoretical framework employed in our calculations. Section~\ref{numerical} is devoted to the presentation and discussion of the numerical results for the magnetic moments. Finally, the main findings of the study are summarized in the concluding section.

\begin{widetext}

\section{Theoretical Formalism}\label{formalism}

The core philosophy of QCD light-cone sum rules is to analyze a correlation function that is sensitive to the property of interest—in this case, the electromagnetic response. This correlation function is calculated in two distinct ways: first, in terms of physical hadronic states (the "hadronic representation"), and second, in terms of QCD degrees of freedom like quarks, gluons, and photon distribution amplitudes (the "QCD representation"). By matching these two representations in a region where they are both valid, and then using a mathematical refinement to isolate the ground state, we can extract the desired hadronic parameter, here the magnetic moment.

The analysis of the electromagnetic structure of the $P_{\psi}^{N}$ states, within the formalism of QCD light-cone sum rules, is initiated through the formulation of the following correlation function:
\begin{equation}
 \label{edmn00}
 \Pi_{\alpha}(p,q) = i^2 \int d^{4}x \int d^{4}y \, e^{ip \cdot x + iq \cdot y} \,
 \langle 0 \big| \mathcal{T} \{ J(x) J_{\alpha}^{\gamma}(y) \bar{J}(0) \} \big| 0 \rangle .
\end{equation}
Here, $J_\alpha^\gamma (y)$ denotes the electromagnetic current, while $J(x)$ is the interpolating current of the $P_{\psi}^{N}$ states, carrying the quantum numbers $J^P = \tfrac{1}{2}^-$.  

This formulation is intuitively clear but can become technically cumbersome, especially when separating short-distance (perturbative) and long-distance (non-perturbative) photon interactions.   
A more elegant and technically advantageous approach is to treat the photon not as a quantum operator but as a classical, weak, external background field. In this picture, the correlation function is defined as:
\begin{equation} \label{edmn01}
\Pi(p,q) = i \int d^4x \, e^{ip \cdot x} \, 
\langle 0 \big| T \{ J(x) \bar{J}(0) \} \big| 0 \rangle_{F},
\end{equation}
where $F$ represents the external background electromagnetic field (EBGEM), with
\begin{equation}
F_{\alpha\beta} = i \big( \varepsilon_\alpha q_\beta - \varepsilon_\beta q_\alpha \big) e^{-iq \cdot x},
\end{equation}
and $\varepsilon_\beta$ and $q_\alpha$ correspond to the photon polarization and four-momentum, respectively.  
It should be emphasized that $\Pi(p,q)$  stands for $\varepsilon^\alpha \Pi_\alpha(p,q)$.  

The EBGEM approach offers a notable advantage by enabling a gauge-invariant separation of soft and hard photon contributions~\cite{Ball:2002ps}. Moreover, the EBGEM is treated as an infinitesimally weak background field, allowing the correlation function in Eq.~(\ref{edmn01}) to be systematically expanded in powers of the field strength:
\begin{equation}
\Pi (p,q) = \Pi^{(0)}(p,q) + \Pi^{(1)}(p,q) + \cdots ,
\end{equation}
where $\Pi^{(0)}(p,q)$ is the correlation function in the absence of the field, which relates to the mass sum rule of the pentaquark (not our focus here), while $\Pi^{(1)}(p,q)$ is the linear response term, representing the interaction with a single photon. This is the term that contains the information about the magnetic moment~\cite{Ball:2002ps,Novikov:1983gd,Ioffe:1983ju}.  
Thus, within the QCD light-cone sum rule framework, the extraction of magnetic and higher multipole moments of the hadrons requires only the computation of the $\Pi^{(1)}(p,q)$ term.

\subsection{The hadronic representation: Parameterizing the physical process}

With the formalism clarified, we turn to the derivation of the QCD light-cone sum rules for the magnetic moments of the $P_{\psi}^{N}$ states. The initial stage of the analysis requires the construction of the correlation function within the hadronic representation. By inserting a complete set of intermediate $P_{\psi}^{N}$ states that share the same quantum numbers as the interpolating currents, the correlation function can be expressed as
 \begin{align}\label{edmn02}
\Pi^{Had}(p,q)&=\frac{\langle0\mid J(x) \mid
{\mathrm{P_{\psi}^{N}}}(p, s) \rangle}{[p^{2}-m_{\mathrm{P_{\psi}^{N}}}^{2}]}
\langle {\mathrm{P_{\psi}^{N}}}(p, s)\mid
{\mathrm{P_{\psi}^{N}}}(p+q, s)\rangle_F 
\frac{\langle {\mathrm{P_{\psi}^{N}}}(p+q, s)\mid
\bar J(0) \mid 0\rangle}{[(p+q)^{2}-m_{\mathrm{P_{\psi}^{N}}}^{2}]}+ \cdots . 
\end{align}

For the subsequent analysis, the hadronic matrix elements appearing in Eq.~(\ref{edmn02}) must be expressed in terms of fundamental hadronic parameters, namely the spinor $u(p,s)$ and the residue $\lambda_{\mathrm{P_{\psi}^{N}}}$, as follows:
\begin{align}
\langle 0 | J(x) | P_{\psi}^{N}(p,s) \rangle &= \lambda_{\mathrm{P_{\psi}^{N}}} \, \gamma_5 \, u(p,s), \label{edmn04} \\
\langle P_{\psi}^{N}(p+q,s) | \bar J(0) | 0 \rangle &= \lambda_{\mathrm{P_{\psi}^{N}}} \, \gamma_5 \, \bar u(p+q,s). \label{edmn004}
\end{align}

For a spin-$\frac{1}{2}$ hadron, the matrix element of the electromagnetic current can be parameterized in terms of Lorentz-invariant form factors, $f_1(q^2)$ and $f_2(q^2)$~\cite{Leinweber:1990dv}:
\begin{align}
\langle P_{\psi}^{N}(p,s)  P_{\psi}^{N}(p+q,s) \rangle_F
&= \varepsilon^\mu \, \bar u(p,s) 
\Big[ \big(f_1(q^2) + f_2(q^2)\big) \gamma_\mu + f_2(q^2) \frac{(2p+q)_\mu}{2 m_{\mathrm{P_{\psi}^{N}}}} \Big] u(p+q,s). \label{edmn005}
\end{align}

By combining Eqs.~(\ref{edmn04})--(\ref{edmn005}) and performing a summation over the spin states, the correlation function in the hadronic representation can be written as
\begin{align}
\label{edmn05}
\Pi^{Had}(p,q)
&= \frac{\lambda_{\mathrm{P_{\psi}^{N}}}^2}{\big[p^2 - m_{P_{\psi}^{N}}^2\big]\big[(p+q)^2 - m_{P_{\psi}^{N}}^2\big]} 
\Big(f_1(q^2)+f_2(q^2)\Big) \Big[ 2 (\varepsilon \cdot p) \pslash 
- m_{\mathrm{P_{\psi}^{N}}} \, \eslash \pslash 
- m_{\mathrm{P_{\psi}^{N}}} \, \eslash \qslash 
+ \pslash \eslash \qslash \Big] .
\end{align}

The magnetic form factor, $F_M(q^2)$, is related to the invariant form factors $f_i(q^2)$ via
\begin{align}
\label{edmn07}
F_M(q^2) = f_1(q^2) + f_2(q^2).
\end{align}
For a real photon, $q^2 = 0$, the magnetic form factor can be directly linked to the magnetic moment of the $P_{\psi}^{N}$ states:
\begin{align}
\label{edmn08}
\mu_{\mathrm{P_{\psi}^{N}}} = \frac{e}{2 m_{\mathrm{P_{\psi}^{N}}}} \, F_M(0).
\end{align}


With these relations, the hadronic side of the correlation function is fully specified, providing a foundation for the subsequent QCD-side evaluation within the light-cone sum rule framework.

\subsection{The QCD representation: Calculating with quarks and gluons}

A critical component in the QCD-side evaluation of the correlation function is the choice of an interpolating current that couples effectively to the pentaquark under study. In constructing such currents, it is crucial to align the quark field arrangements with both the valence quark content and the quantum numbers of the hidden-charm pentaquark states. The one-gluon exchange interaction favors the formation of diquarks in the color antitriplet channel, rendering certain diquark configurations more stable. Previous QCD sum rule studies~\cite{Wang:2010sh, Kleiv:2013dta} suggest that scalar and axial-vector diquark correlations provide the most favorable building blocks for constructing pentaquark interpolating currents. 
Guided by these considerations, in the present work we employ interpolating currents of the axial-vector–diquark–scalar–diquark–antiquark type as well as the axial-vector–diquark–axial-vector–diquark–antiquark type. Assuming that the pentaquark under investigation possesses quantum numbers $J^P = \frac{1}{2}^-$, the corresponding interpolating currents can be explicitly written as follows~\cite{Wang:2025qtm}: 
\begin{align}\label{curpcs2}
J_{1}(x)&= \varepsilon^{abc}\varepsilon^{ade} \varepsilon^{bfg}\bigg\{ \big[ {u}^T_d(x) C \gamma_5 {d}_e(x) \big] \big[ {u}^T_f(x) C \gamma_5 c_g(x)\big]   \bigg\}  C  \bar{c}^{T}_{c}(x) \, , \\
J_{2}(x)&= \varepsilon^{abc}\varepsilon^{ade} \varepsilon^{bfg}\bigg\{ \big[ {u}^T_d(x) C \gamma_5 {d}_e(x) \big] \big[ {u}^T_f(x) C \gamma_\mu c_g(x)\big]   \bigg\}   \gamma_5 \gamma^\mu C  \bar{c}^{T}_{c}(x) \, , \\
J_{3}(x)&=\frac{\varepsilon^{abc}\varepsilon^{ade} \varepsilon^{bfg}}{\sqrt{2}} \bigg\{ \big[ {u}^T_d(x) C \gamma_\mu {u}_e(x) \big] \big[ {d}^T_f(x) C \gamma_5 c_g(x)\big]   - 
\big[ {u}^T_d(x) C \gamma_\mu {d}_e(x) \big] \big[ {u}^T_f(x) C \gamma_5c_g(x)\big]  \bigg\}  \gamma_5 \gamma^\mu C  \bar{c}^{T}_{c}(x) \, , \\
J_{4}(x)&=\frac{\varepsilon^{abc}\varepsilon^{ade} \varepsilon^{bfg}}{\sqrt{2}} \bigg\{ \big[ {u}^T_d(x) C \gamma_\mu {u}_e(x) \big] \big[ {d}^T_f(x) C \gamma^\mu c_g(x)\big]   - 
\big[ {u}^T_d(x) C \gamma_\mu {d}_e(x) \big] \big[ {u}^T_f(x) C \gamma^\mu c_g(x)\big]  \bigg\}  C  \bar{c}^{T}_{c}(x) \, , 
\end{align}
where $a$, $b$, $\cdots$ are color indexes and the $C$ is the charge conjugation operator.
The four interpolating currents $J_1(x),\dots,J_4(x)$ used in this work have been studied in detail in Ref.~\cite{Wang:2025qtm} and are all assigned the spin–parity $J^{P} = \frac{1}{2}^{-}$. They share the same quark content ($uud\bar{c}c$) but differ in their internal diquark organization: $J_1(x)$ contains two scalar diquarks $[ud]_{0^{+}}$ and $[uc]_{0^{+}}$; $J_2(x)$ consists of a scalar light diquark $[ud]_{0^{+}}$ and an axial-vector heavy diquark $[uc]_{1^{+}}$; $J_3(x)$ is built from an axial-vector light diquark ($[uu]_{1^{+}}$ or $[ud]_{1^{+}}$) combined with a scalar heavy diquark ($[dc]_{0^{+}}$ or $[uc]_{0^{+}}$); and $J_4(x)$ contains two axial-vector diquarks $[uu]_{1^{+}}$ (or $[ud]_{1^{+}}$) and $[dc]_{1^{+}}$ (or $[uc]_{1^{+}}$). Following the classification of Ref.~\cite{Wang:2025qtm}, $J_1(x)$ and $J_3(x)$ are predominantly scalar-diquark type, whereas $J_2(x)$ and $J_4(x)$ are predominantly axial-vector-diquark type. We therefore expect their magnetic moments to exhibit distinct patterns due to this differing diquark composition.  
%


The computation of the correlation function on the QCD side, $\Pi^{QCD}(p,q)$, is performed in the deep Euclidean region, where $p^2 \ll 0$ and $(p+q)^2 \ll 0$, to ensure the convergence of the operator product expansion (OPE). The analysis proceeds according to a systematic workflow for each interpolating current $J_i(x)$: we begin by inserting the explicit form of the current into the correlation function; subsequently, Wick's theorem is employed to contract all pairs of quark fields. These manipulations lead to the following generic form for the correlation function associated with each current:
\begin{align} 
\Pi^{QCD-J_1(x)}(p,q)&= -i\,\varepsilon^{abc}\varepsilon^{a^{\prime}b^{\prime}c^{\prime}}\varepsilon^{ade} \varepsilon^{a^{\prime}d^{\prime}e^{\prime}}\varepsilon^{bfg} \varepsilon^{b^{\prime}f^{\prime}g^{\prime}} 
\int d^4x e^{ip\cdot x}
\langle 0|
\Big\{
 \mbox{Tr}\Big[  \gamma_5 S_{d}^{ee^\prime}(x) \gamma_5 C   S_{u}^{dd^\prime \mathrm{T}}(x) C\Big]
\nonumber\\
& \times 
  \mbox{Tr}\Big[ \gamma_5 S_c^{gg^\prime}(x) \gamma_5 C  S_{u}^{ff^\prime \mathrm{T}}(x)C \Big] 
\Big \} 
\Big(  C S_c^{c^{\prime}c \mathrm{T}} (-x) C  \Big)
|0 \rangle_F ,  \label{QCD1}\\
\Pi^{QCD-J_2(x)}(p,q)&= -i\,\varepsilon^{abc}\varepsilon^{a^{\prime}b^{\prime}c^{\prime}}\varepsilon^{ade} \varepsilon^{a^{\prime}d^{\prime}e^{\prime}}\varepsilon^{bfg} \varepsilon^{b^{\prime}f^{\prime}g^{\prime}} 
\int d^4x e^{ip\cdot x}
\langle 0|
\Big\{
 \mbox{Tr}\Big[  \gamma_5 S_{d}^{ee^\prime}(x) \gamma_5 C   S_{u}^{dd^\prime \mathrm{T}}(x) C\Big]
\nonumber\\
& \times 
  \mbox{Tr}\Big[ \gamma_\mu S_c^{gg^\prime}(x) \gamma_\nu C  S_{u}^{ff^\prime \mathrm{T}}(x)C \Big] 
\Big \} 
\Big( \gamma_5 \gamma^\mu C S_c^{c^{\prime}c \mathrm{T}} (-x) C \gamma^\nu   \gamma_5 \Big)
|0 \rangle_F ,  \label{QCD2}
\end{align}
\begin{align} 
\Pi^{QCD-J_3(x)}(p,q)&= -\frac{i}{2}\,\varepsilon^{abc}\varepsilon^{a^{\prime}b^{\prime}c^{\prime}}\varepsilon^{ade} \varepsilon^{a^{\prime}d^{\prime}e^{\prime}}\varepsilon^{bfg} \varepsilon^{b^{\prime}f^{\prime}g^{\prime}} 
\int d^4x e^{ip\cdot x}
\langle 0|
\Big\{
\nonumber\\
&
  \mbox{Tr}\Big[  \gamma_\mu S_{u}^{ee^\prime}(x) \gamma_\nu C   S_{u}^{dd^\prime \mathrm{T}}(x) C\Big]
 \mbox{Tr}\Big[ \gamma^\mu S_c^{gg^\prime}(x) \gamma^\nu C  S_{d}^{ff^\prime \mathrm{T}}(x)C \Big] 
 \nonumber\\
&
- \mbox{Tr}\Big[  \gamma_\mu S_{u}^{ed^\prime}(x) \gamma_\nu C  S_{u}^{de^\prime \mathrm{T}}(x) C\Big]
 \mbox{Tr}\Big[ \gamma^\mu S_c^{gg^\prime}(x) \gamma^\nu C  S_{d}^{ff^\prime \mathrm{T}}(x)C  \Big] 
\nonumber\\
&
  + \mbox{Tr}\Big[  \gamma_\mu S_{d}^{ee^\prime}(x) \gamma_\nu C   S_{u}^{ff^\prime \mathrm{T}}(x) C\Big]
 \mbox{Tr}\Big[ \gamma^\mu S_c^{gg^\prime}(x) \gamma^\nu C  S_{u}^{ff^\prime \mathrm{T}}(x)C \Big]
\Big \} 
\Big( C S_c^{c^{\prime}c \mathrm{T}} (-x) C  \Big)
|0 \rangle_F, \label{QCD3} 
\\
\Pi^{QCD-J_4(x)}(p,q)&= -\frac{i}{2}\,\varepsilon^{abc}\varepsilon^{a^{\prime}b^{\prime}c^{\prime}}\varepsilon^{ade} \varepsilon^{a^{\prime}d^{\prime}e^{\prime}}\varepsilon^{bfg} \varepsilon^{b^{\prime}f^{\prime}g^{\prime}} 
\int d^4x e^{ip\cdot x}
\langle 0|
\Big\{
\nonumber\\
&
  \mbox{Tr}\Big[  \gamma_\mu S_{u}^{ee^\prime}(x) \gamma_\nu C   S_{u}^{dd^\prime \mathrm{T}}(x) C\Big]
 \mbox{Tr}\Big[ \gamma_5 S_c^{gg^\prime}(x) \gamma_5 C  S_{d}^{ff^\prime \mathrm{T}}(x)C \Big] 
 \nonumber\\
&
- \mbox{Tr}\Big[  \gamma_\mu S_{u}^{ed^\prime}(x) \gamma_\nu C  S_{u}^{de^\prime \mathrm{T}}(x) C\Big]
 \mbox{Tr}\Big[ \gamma_5 S_c^{gg^\prime}(x) \gamma_5 C  S_{d}^{ff^\prime \mathrm{T}}(x)C  \Big] 
\nonumber\\
&
  + \mbox{Tr}\Big[  \gamma_\mu S_{d}^{ee^\prime}(x) \gamma_\nu C   S_{u}^{ff^\prime \mathrm{T}}(x) C\Big]
 \mbox{Tr}\Big[ \gamma_5 S_c^{gg^\prime}(x) \gamma_5 C  S_{u}^{ff^\prime \mathrm{T}}(x)C \Big]
\Big \} 
\Big( \gamma_5 \gamma^\mu C S_c^{c^{\prime}c \mathrm{T}} (-x) C \gamma^\nu   \gamma_5 \Big)
|0 \rangle_F, \label{QCD4}
\end{align}
where the functions $\Pi^{QCD-J_i(x)}(p,q)$ encapsulate the results of the Wick contractions and will be evaluated using the quark propagators and photon DAs. The explicit forms of the light and charm quark propagators employed in this calculation are given by~\cite{Balitsky:1987bk, Belyaev:1985wza}:
\begin{align}
\label{edmn13}
S_{q}(x)&= S_q^{free}(x) 
-i\frac { g_s }{16 \pi^2 x^2} \int_0^1 du \, G^{\mu \nu} (ux)
\bigg[\bar u \rlap/{x} 
\sigma_{\mu \nu} + u \sigma_{\mu \nu} \rlap/{x}
 \bigg],\\
%
S_{c}(x)&=S_c^{free}(x)
-i\frac{m_{c}\,g_{s} }{16\pi ^{2}}  \int_0^1 du \,G^{\mu \nu}(ux)\bigg[ (\sigma _{\mu \nu }{\xslash}
+{\xslash}\sigma _{\mu \nu }) 
    \frac{K_{1}\big( m_{c}\sqrt{-x^{2}}\big) }{\sqrt{-x^{2}}}
 +2\sigma_{\mu \nu }K_{0}\big( m_{c}\sqrt{-x^{2}}\big)\bigg],
 \label{edmn14}
\end{align}%
with  
\begin{align}
 S_q^{free}(x)&=\frac{1}{2 \pi x^2}\Big(i \frac{\xslash}{x^2}- \frac{m_q}{2}\Big),\\
 S_c^{free}(x)&=\frac{m_{c}^{2}}{4 \pi^{2}} \bigg[ \frac{K_{1}\big(m_{c}\sqrt{-x^{2}}\big) }{\sqrt{-x^{2}}}
+i\frac{{\xslash}~K_{2}\big( m_{c}\sqrt{-x^{2}}\big)}
{(\sqrt{-x^{2}})^{2}}\bigg].
\end{align}
In this context, $ G^{\mu\nu} (x) $ represents the background gluonic field strength tensor, while $ K_0\big( m_c \sqrt{-x^2} \big) $, $ K_1\big( m_c \sqrt{-x^2} \big) $, and $ K_2\big( m_c \sqrt{-x^2} \big) $ denote the modified Bessel functions of the second kind.

The correlation function receives contributions from two distinct physical mechanisms, classified by the distance scale of the photon-quark interaction: the perturbative (short-distance) and non-perturbative (long-distance) components. A complete description mandates the inclusion of both, as they are complementary in mapping the electromagnetic structure of the hadron. 
The perturbative component corresponds to the process where the photon is absorbed in a hard scattering event, directly coupling to a quark at a short distance. In our computational scheme, this contribution is calculated by considering the photon interaction with \textit{free} quark propagators. This approach captures the leading-order hard interaction while providing a technically manageable framework. Mathematically, for the perturbative insertion on a given quark line, we implement the substitution:
\begin{align}
\label{free}
S^{\mathrm{free}}(x) \rightarrow \int d^4y \, S^{\mathrm{free}}(x-y) \,\rlap/{\!A}(y) \, S^{\mathrm{free}}(y).
\end{align}
 In the evaluation of the correlation functions in Eqs.~(\ref{QCD1})--(\ref{QCD4}), the full propagators including non-perturbative gluonic components are used as the foundation. However, when calculating the specific perturbative photon contributions, we employ the substitution in Eq.~(\ref{free}) while setting all propagators in that particular diagram to their free forms, thus isolating the hard photon interaction from the non-perturbative background. This defines a consistent perturbative expansion where the photon interacts with quark fields that are not dressed by gluonic corrections.
 
In contrast, the non-perturbative component arises from the soft interaction of the photon with the QCD vacuum condensates. Here, the photon fluctuates into a quark-antiquark pair that engages in long-distance interactions with the vacuum fields. This process is parameterized by the photon distribution amplitudes (DAs) and is implemented by replacing a light-quark propagator with its non-local matrix element:
\begin{align}
\label{edmn21}
S_{\alpha\beta}^{ab}(x) \rightarrow -\frac{1}{4} \langle \gamma(q) | \bar{q}^a(x) \Gamma_i q^b(0) | 0 \rangle (\Gamma_i)_{\alpha\beta},
\end{align}
where $\Gamma_i = \{ \mathbb{1}, \gamma_5, \gamma_\mu, i\gamma_5 \gamma_\mu, \sigma_{\mu\nu}/2 \}$ spans the complete set of Dirac structures. The resulting matrix elements $\langle \gamma(q) | \bar{q}(x) \Gamma_i q(0) | 0 \rangle$ and $\langle \gamma(q) | \bar{q}(x) \Gamma_i G_{\alpha\beta} q(0) | 0 \rangle$ define the photon DAs of increasing twist~\cite{Ball:2002ps}, which we employ up to twist-4 accuracy. 
Within this scheme, heavy quarks are treated exclusively through the perturbative approach described above. For the charm quark, while its propagator $S_c(x)$ does contain non-perturbative gluonic corrections, its interaction with the photon is treated perturbatively in the sense that we do not consider non-perturbative photon DAs  contributions involving charm quarks. The $m_c \gg \Lambda_{QCD}$ hierarchy ensures that such long-distance photon-charm interactions are suppressed. 

The full QCD representation $\Pi^{\mathrm{QCD}}(p,q)$ for each current $J_i(x)$ is constructed by systematically evaluating all diagrams where the photon is inserted perturbatively on every possible quark line (using free propagators) and non-perturbatively on each light-quark line (using the DA replacement). This comprehensive procedure involves extensive computation of integrals using light-cone expansion techniques and Fourier transformations to momentum space.

  \subsection{Deriving the sum rule and isolating the magnetic moment}

We now have two expressions for the same correlation function: $\Pi^{Had}(p,q)$, written in terms of the magnetic moment $\mu$, mass $m$, and coupling $\lambda$; and $\Pi^{QCD}(p,q)$, a complicated function of QCD parameters, photon DAs, and kinematic variables. 
The QCD sum rule is established by equating these two representations:
\begin{equation}
\Pi^{Had}(p,q) \equiv \Pi^{QCD}(p,q).
\end{equation}  

However, both representations contain contributions beyond the ground state. The hadronic representation includes excited states and the continuum explicitly, which we initially denoted by an ellipsis. On the QCD side, the OPE  also incorporates effects from all energy scales, including those corresponding to excited hadronic states.  To isolate the ground-state contribution, we invoke the principle of quark-hadron duality. This principle posits that above an effective continuum threshold $\mathrm{s_0}$, the spectral density of the hadronic continuum is equivalent to the imaginary part of the QCD correlator. This allows us to subtract the continuum contributions from both sides, leaving only the ground-state terms. 
To further suppress any residual continuum contributions and enhance the convergence of the OPE, we apply a Borel transformation to both sides of the matched equation. This mathematical operation, parameterized by the Borel mass $\mathrm{M^2}$, exponentially suppresses higher-energy states. 
After applying the Borel transformation and continuum subtraction, we match the coefficients of the same Lorentz structures from both sides. This procedure yields the final sum rules for the magnetic moments. For each current $J_i$, the sum rule takes the generic form:
\begin{equation}
\mu^{J_i}_{P_{\psi}^{N}} \,  \lambda^{2,J_i}_{P_{\psi}^{N}} \, e^{-m^2 / \mathrm{M^2}} = \mathcal{B} \left\{ \Pi^{QCD}_{J_i} \right\}(\mathrm{M^2},\mathrm{s_0}),
\end{equation}
where $\mathcal{B}$ denotes the Borel-transformed and continuum-subtracted QCD expression. More concretely, we can write:
\begin{align}
\label{edmn15}
\mu_{J_i} \, \lambda_{J_i}^2  = e^{-m_{J_i}^2/\mathrm{M^2}}\,\rho_i(\mathrm{M^2}, \mathrm{s_0}), \quad i = 1,2,3,4
\end{align}
where the index $i$ labels the four interpolating currents.  
The explicit expressions for the functions $\rho_i (\mathrm{M^2}, s_0)$ derived following the full implementation of the procedures described above, are provided in the Appendix.  They encapsulate all the perturbative and non-perturbative contributions described in the previous subsection, after the Borel transformation and continuum subtraction have been applied.

This concludes the formal derivation of the QCD light-cone sum rules for the magnetic moments of the $P_{\psi}^{N}$ pentaquark states. The subsequent numerical analysis will rely entirely on the framework established here.
\end{widetext}

\section{Results and discussions}\label{numerical}

This section focuses on the analysis of the sum rules derived for the electromagnetic properties of the considered states. The numerical values of all input parameters employed in these calculations are listed in Table~\ref{inputparameter}. For the subsequent evaluations, the photon DAs and their corresponding input parameters have been adopted from the results compiled in Ref.~\cite{Ball:2002ps}.
 \begin{table}[htb!]
	\addtolength{\tabcolsep}{10pt}
	\caption{Values of the input parameters used in the numerical evaluation of the sum rules for the electromagnetic properties.}
	\label{inputparameter}
\begin{tabular}{l|ccc}
               \hline\hline
                \\
Inputs & Values \\
 \\
                                        \hline\hline
$m_c$&$ 1.273 \pm 0.0046$~GeV \cite{ParticleDataGroup:2024cfk}\\
$m_{P_{\psi}^{N}}^{J_1(x)}$ & $4.31 \pm 0.11$~GeV \cite{Wang:2025qtm} \\
$m_{P_{\psi}^{N}}^{J_2(x)}$ & $4.45 \pm 0.11$~GeV \cite{Wang:2025qtm} \\
$m_{P_{\psi}^{N}}^{J_3(x)}$ & $4.20 \pm 0.11$~GeV \cite{Wang:2025qtm} \\
$m_{P_{\psi}^{N}}^{J_4(x)}$ & $4.25 \pm 0.11$~GeV \cite{Wang:2025qtm} \\
$\lambda_{P_{\psi}^{N}}^{J_1(x)} $&$ (1.40 \pm 0.23)\times 10^{-3} $~GeV$^6$ \cite{Wang:2025qtm}\\
$ \lambda_{P_{\psi}^{N}}^{J_2(x)} $&$(3.02 \pm 0.48)\times 10^{-3} $~GeV$^6$ \cite{Wang:2025qtm}\\
$\lambda_{P_{\psi}^{N}}^{J_3(x)} $&$ (2.24 \pm 0.40)\times 10^{-3} $~GeV$^6$ \cite{Wang:2025qtm}\\
$ \lambda_{P_{\psi}^{N}}^{J_4(x)} $&$(2.78 \pm 0.47)\times 10^{-3} $~GeV$^6$ \cite{Wang:2025qtm}\\
$\langle \bar qq\rangle $&$ (-0.24 \pm 0.01)^3 $~GeV$^3$ \cite{Ioffe:2005ym}\\
$ \langle g_s^2G^2\rangle  $&$ 0.48 \pm 0.14 $~GeV$^4$ \cite{Narison:2018nbv}\\
$f_{3\gamma} $&$ -0.0039 $~GeV$^2$ \cite{Ball:2002ps}\\
$\chi $&$ -2.85 \pm 0.5 $~GeV$^{-2}$ \cite{Rohrwild:2007yt}\\
                                      \hline\hline
 \end{tabular}
\end{table}


In addition to the previously listed input parameters, the analysis involves two auxiliary quantities: the Borel mass parameter, $\mathrm{M^2}$, and the continuum threshold, $\mathrm{s_0}$. These parameters are determined by employing the standard procedures and stability requirements inherent to the QCD sum rule framework. The acceptable range for $\mathrm{M^2}$ is established by ensuring the proper convergence of the OPE, referred to as CVG, and by maintaining the dominance of the ground-state pole contribution (PC) over the continuum. Specifically, the lower bound of $\mathrm{M^2}$ is set by requiring the OPE series to converge adequately, keeping the CVG sufficiently small. Conversely, the upper bound is constrained by maximizing the pole contribution to justify the assumption of single-pole dominance. By imposing these two conditions, the Borel window for $\mathrm{M^2}$ can be determined, and an optimal value of the continuum threshold $\mathrm{s_0}$ can be obtained by minimizing the dependence of the magnetic moments of the $P_{\psi}^{N}$ states on the Borel parameter $\mathrm{M^2}$. These conditions can be formally expressed as
\begin{align}
 \text{PC} &= \frac{\rho_i (\mathrm{M^2},\mathrm{s_0})}{\rho_i (\mathrm{M^2},\infty)} \geq 40\%, \\
 \text{CVG}  &= \frac{\rho_i^{\text{DimN}} (\mathrm{M^2},\mathrm{s_0})}{\rho_i (\mathrm{M^2},\mathrm{s_0})} < 5\%,
\end{align}
where $\rho_i^{\text{DimN}} (\mathrm{M^2},s_0)$ denotes the highest-dimensional terms in the operator product expansion of $\rho_i (\mathrm{M^2},s_0)$. On the QCD side, the highest-order contributions arise from dimension-7 operators, specifically $\langle g_s^2 G^2 \rangle \langle \bar q q \rangle$ (D7). Consequently, the CVG analysis has been performed by including these D7 terms, and the numerical results presented in Table~\ref{parameter} reflect this choice.  Once these conditions were verified, we proceeded with confidence regarding the reliability of our predictions. Moreover, the QCD representation includes a range of operator structures, such as $\langle g_s^2 G^2 \rangle f_{3\gamma}$ (D6), $\langle g_s^2 G^2 \rangle \langle \bar q q \rangle \chi$ (D5), $\langle \bar q q \rangle$ (D3), $f_{3\gamma}$ (D2), and $\langle \bar q q \rangle \chi$ (D1).  As an illustration, Figure~\ref{Msqfig} shows that, within the chosen Borel window, the PC  significantly exceeds the continuum contribution, thereby confirming the dominance of the ground state. At the same time, the relative sizes of the condensate terms indicate a satisfactory convergence of the OPE series. The analysis reveals that, within the CVG framework, the dominant non-perturbative effect arises from the term proportional to D1. This is followed, in decreasing order of significance, by the contributions of D2, D3, D5, D6, and D7. 
For completeness, Fig.~\ref{Msqfig1} also depicts the dependence of the magnetic dipole moment of the $P_{\psi}^{N}$ pentaquark on both the Borel mass parameter $\mathrm{M^2}$ and the continuum threshold $\mathrm{s_0}$. As expected, only minor variations are observed within the selected parameter ranges, although a residual sensitivity to these parameters persists and constitutes a source of uncertainty.
\begin{widetext}

    \begin{figure}[htb!]
\includegraphics[width=0.4\textwidth]{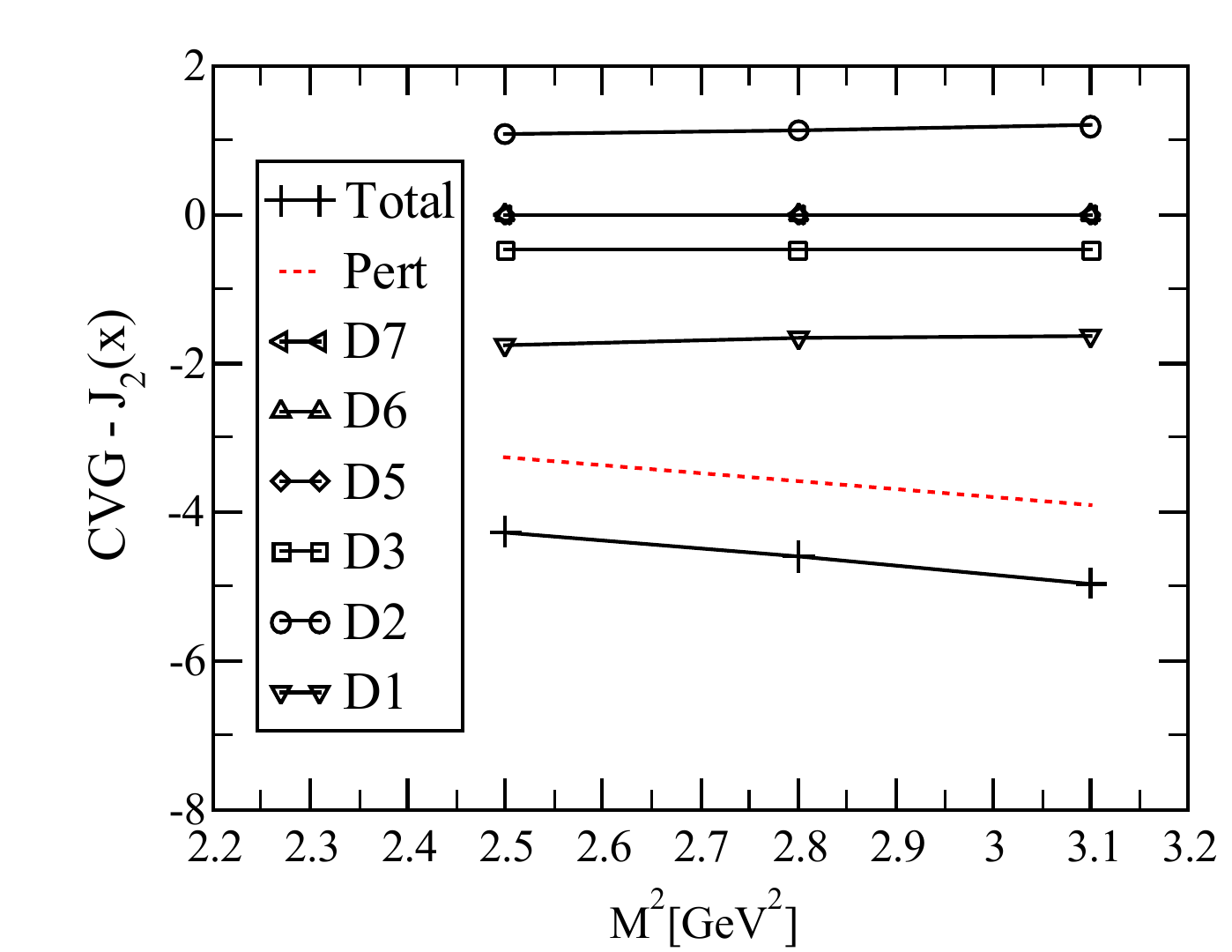}~~~~
\includegraphics[width=0.4\textwidth]{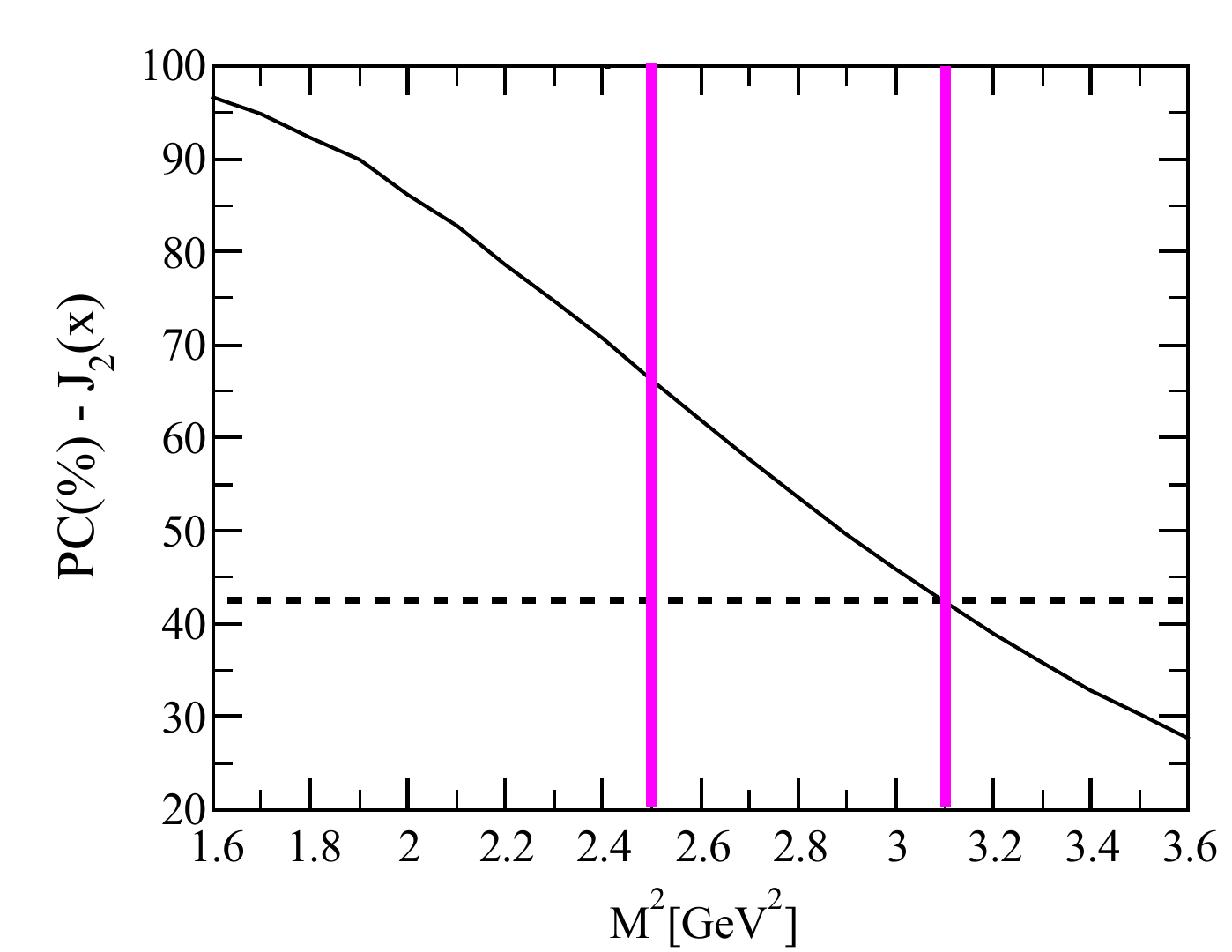}~~\\
\caption{CVG (left panel) and PC (right panel) analyses for the magnetic dipole moment of the $P_{\psi}^{N}$ pentaquark, obtained from the current $J_2(x)$ as a function of the Borel parameter $\mathrm{M}^{2}$. The curves are computed using $\mathrm{s}_{0}$-averaging over the continuum threshold interval given in Table~\ref{parameter}. In the right panel, the vertical lines mark the chosen Borel region, and the horizontal line indicates the minimum $\mathrm{s}_{0}$-averaged PC value within this region.}
 \label{Msqfig}
  \end{figure}
  
 
\begin{figure}[htb!]
\includegraphics[width=0.4\textwidth]{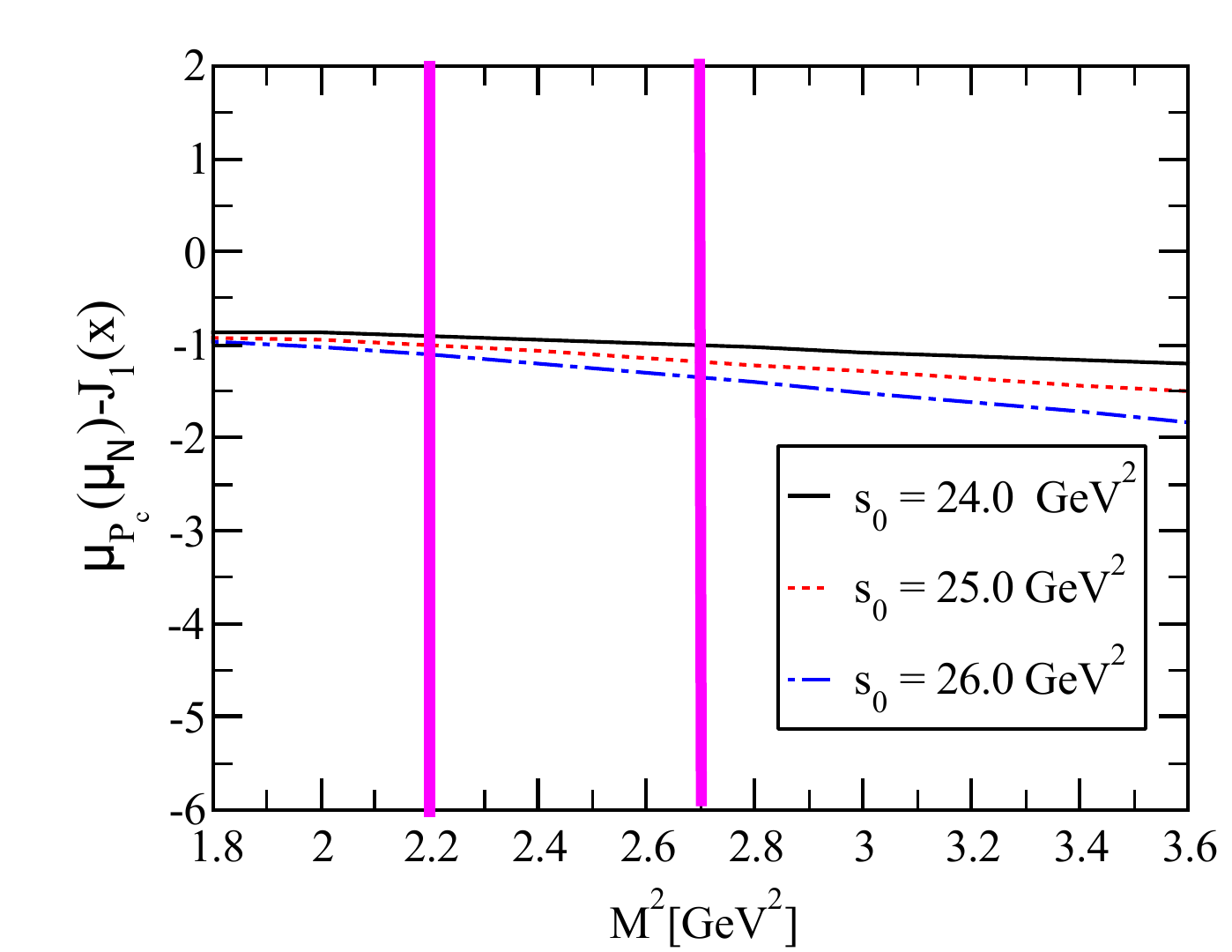}~~~~
\includegraphics[width=0.4\textwidth]{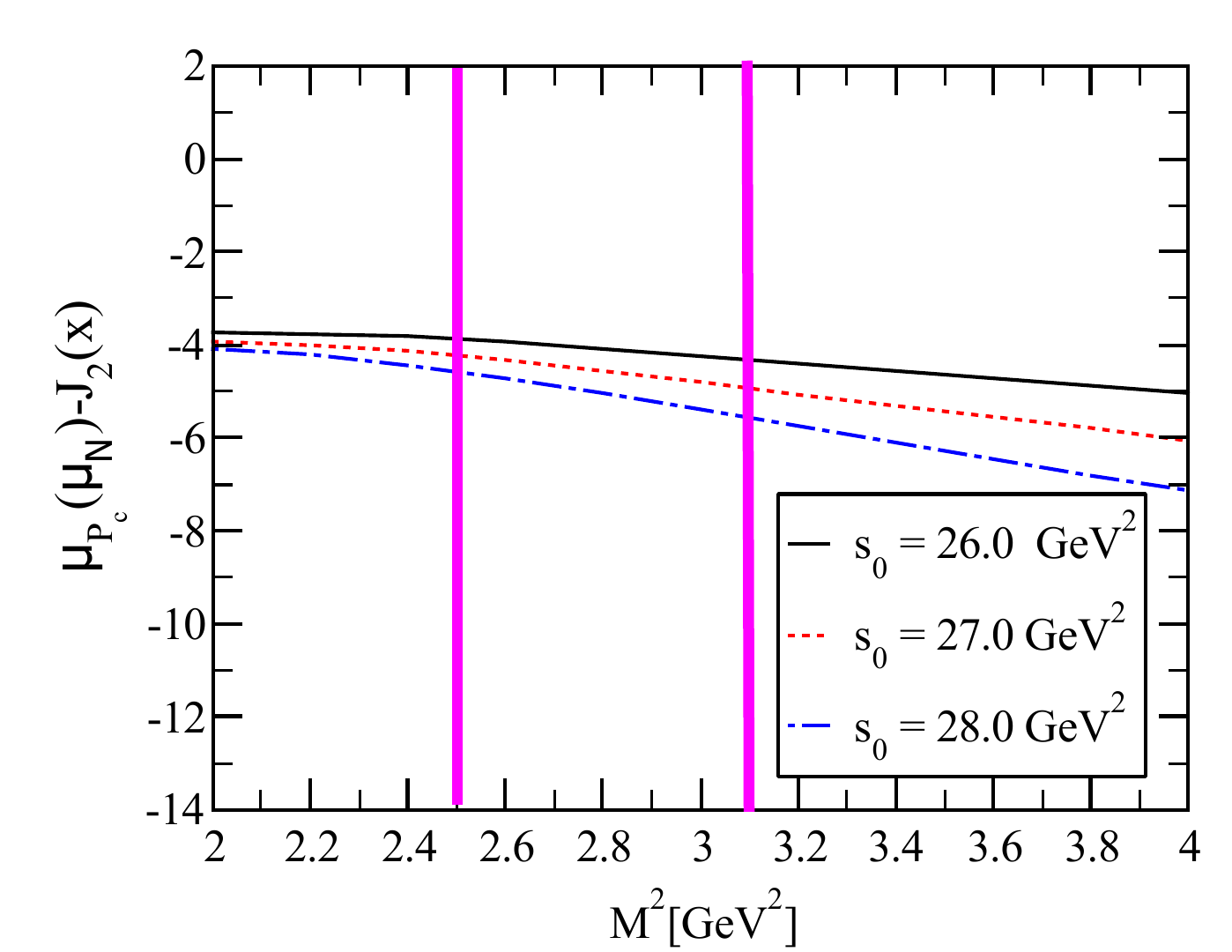}~~\\
\includegraphics[width=0.4\textwidth]{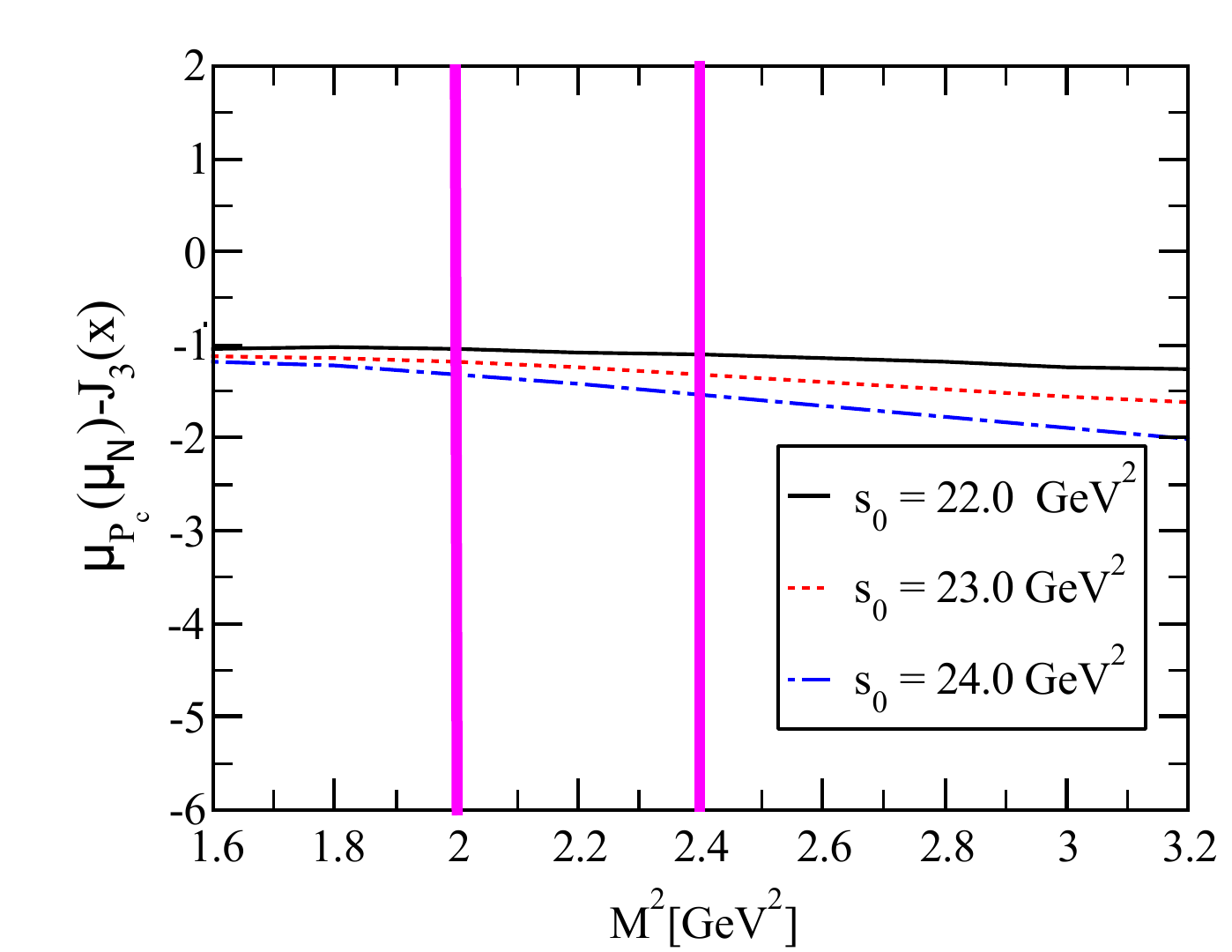}~~~~
\includegraphics[width=0.4\textwidth]{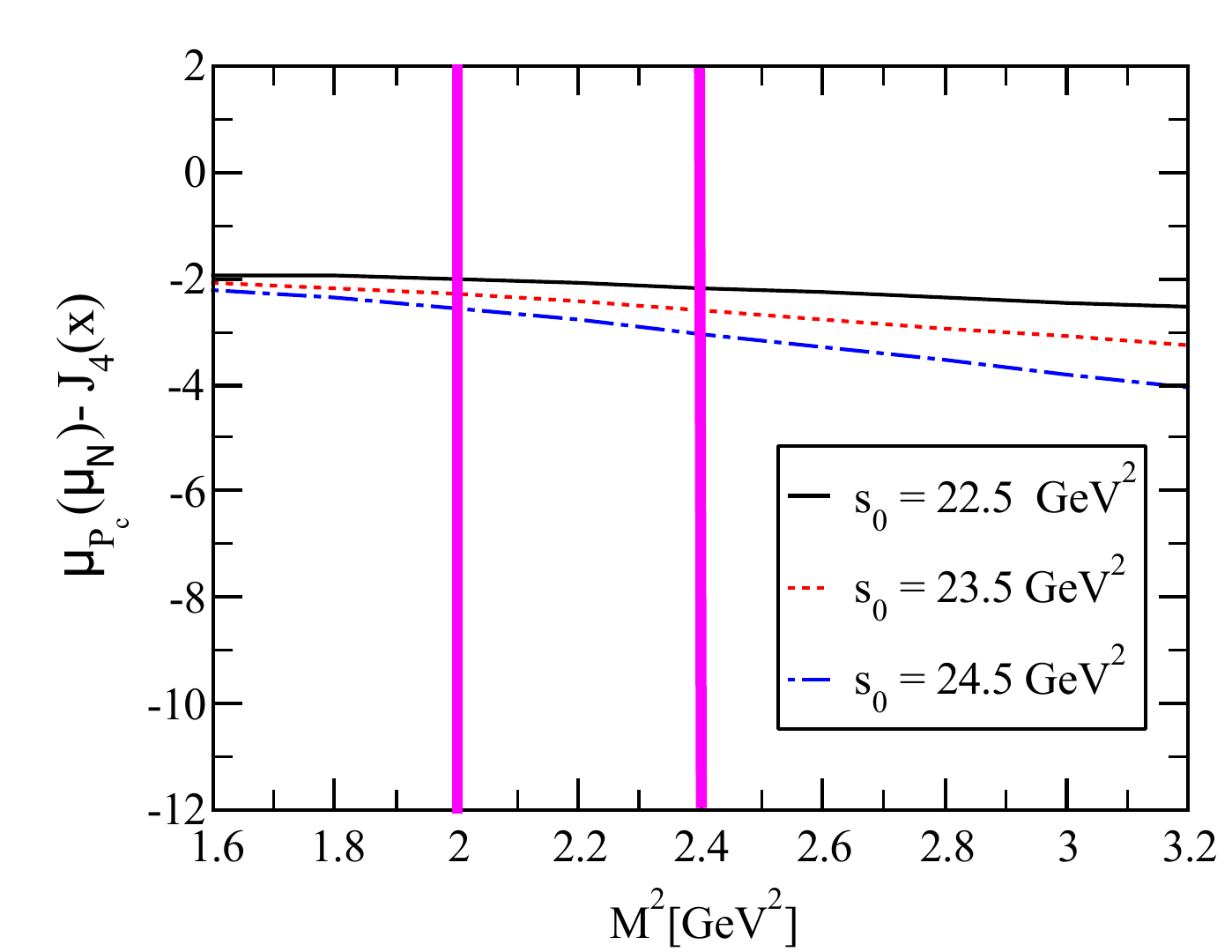}\\
\caption{Magnetic moments of the $P_{\psi}^{N}$ pentaquarks as a function of $\mathrm{M^2}$ for three different values of $\mathrm{s_0}$. The region enclosed between the vertical lines denotes the working interval of $\mathrm{M^2}$.}
 \label{Msqfig1}
  \end{figure}
  
 \end{widetext} 

\begin{widetext} 

\begin{table}[htb!]
	\addtolength{\tabcolsep}{10pt}
	\caption{	The working intervals of $\mathrm{s_0}$ and $\mathrm{M^2}$, determined by the CVG and PC criteria for the magnetic moment calculations of the $P_{\psi}^{N}$ pentaquarks, together with the numerical values of the magnetic dipole moment.}
	\label{parameter}
	\begin{ruledtabular}
\begin{tabular}{l|cccccc}
               \\
States &$\mathrm{s_0}$ (GeV$^2$) & $\mathrm{M^2}$ (GeV$^2$)&~~PC ($\%$)~~&~~ CVG ($\%$) & $\mu_{P_{\psi}^{N}} (\mu_N)$\\
         \\
                                 \hline\hline
$J_1(x)$ & $24.0-26.0$ & $2.20-2.70$ & $66.47-44.02$ &  $ < 1$  & $-1.10^{+0.26}_{-0.22}$
                      \\
$J_2(x)$ & $26.0-28.0$ & $2.50-3.10$ & $66.19-42.42$ &  $< 1$  & $ -4.59^{+0.98}_{-0.75}$ 
                       \\
$J_3(x)$ & $22.0-24.0$ & $2.00-2.40$ & $64.35-45.04$ &  $ < 1$  & $-1.25^{+0.30}_{-0.26}$
                        \\
$J_4(x)$ & $22.5-24.5$ & $2.00-2.40$ & $64.02-43.72$ &  $< 1$  & $-2.44^{+0.62}_{-0.45}$ 
                          \\
 \end{tabular}
\end{ruledtabular}
\end{table}

\end{widetext}
  
The final results for the magnetic moments are summarized in Table~\ref{parameter}. The quoted uncertainties reflect variations in the input parameters, as well as in the auxiliary quantities, such as $\mathrm{s_0}$ and $\mathrm{M^2}$, and in the parameters associated with the DAs. Our findings indicate that the magnetic moments corresponding to both interpolating currents of the $P_{\psi}^{N}$ pentaquarks lie within ranges that may be accessible in future experimental measurements. Interestingly, although the four interpolating currents share the same quark content and identical quantum numbers, their corresponding magnetic moments differ significantly, highlighting the sensitivity of this observable to variations in the internal structure, particularly the configuration of diquarks within the hadrons. These results indicate that magnetic moments are strongly influenced by both the composition and spin alignment of the constituent diquarks. As a result, electromagnetic observables such as magnetic moments provide a powerful probe of the internal quark--gluon dynamics and can help distinguish between alternative structural arrangements in hadrons. In experimental contexts, such insights may aid in determining quantum numbers and in discriminating among competing theoretical models of hadron structure. Although direct experimental determination of the magnetic moments of short-lived hadrons remains challenging, theoretical analyses based on QCD-inspired frameworks continue to provide important insights into their internal structure and dynamics. A more comprehensive understanding could be obtained by complementing the present study with investigations of the decay channels, branching ratios, and additional electromagnetic properties of the $P_{\psi}^{N}$ states.

Ordinarily, one could anticipate that changes in the chosen basis, such as variations in charge, spin, or isospin, would leave the theoretical outcomes largely unaffected. For magnetic moments, however, this assumption breaks down. The reason is that electromagnetic observables are directly connected to the internal configuration of the hadron.    
In the context of electromagnetic properties, any adjustment to the basis of the considered state inherently affects the hadron’s internal structure, which can in turn produce appreciable differences in the predicted outcomes. 
Previous studies have demonstrated that the magnetic moments of multiquark systems are particularly sensitive to such changes, displaying pronounced differences under alternative structural assumptions (see Refs.~\cite{Ozdem:2025jda, Wang:2016dzu, Ozdem:2024rch, Gao:2021hmv, Li:2024wxr, Li:2024jlq, Ozdem:2024rqx, Ozdem:2024txt, Ozdem:2024dbq, Ozdem:2024lpk, Azizi:2023gzv}).  This indicates that the selection of interpolating currents and wave functions—equivalently, the specification of isospin, spin, and charge—has a decisive influence on the obtained results. 

Table \ref{table3} presents the magnetic dipole contributions of light ($u, d$) and charm ($c$) quarks for the $P_{\psi}^{N}$ pentaquark states, along with the relative ratios of these contributions. In order to quantify the contributions of individual quark flavors, the charge factors ($e_q$ and $e_c$) within the sum rules are deliberately manipulated. Specifically, the contribution from the light quarks is isolated by setting $e_c = 0$ in Eqs.~(\ref{F1sonuc})–(\ref{F4sonuc}). This procedure eliminates all terms proportional to the charm quark charge, ensuring that only those terms associated with $e_q$ remain, thereby allowing a clear separation of the light quark sector from the total magnetic moment.  We observe that the internal structure of the interpolating currents strongly governs the distribution of quark contributions. Specifically, scalar-heavy diquark currents ($J_1(x)$ and $J_3(x)$) contain a scalar-light diquark ($[ud]$) coupled to a scalar ($J_1(x)$) or scalar-light–vector-heavy diquark ($J_3(x)$) involving the charm quark. In both cases, the charm quark dominates the total magnetic moment, contributing approximately 98–99\% of $\mu_{tot}$. This is because the charm quark is bound in a relatively compact heavy-light diquark, which aligns its spin coherently with the total spin of the pentaquark, leading to a large constructive contribution. The light quark contributions in these currents are strongly suppressed due to spin anti-alignment and smaller magnetic moments of light quarks. Vector-heavy diquark and symmetric difference currents ($J_2(x)$ and $J_4(x)$) contain a scalar-light diquark coupled to a vector-heavy diquark ($J_2(x)$), or vector diquarks for both light and heavy sectors with an antisymmetric combination ($J_4(x)$). These structural differences enhance the contributions of the light quarks. In $J_2(x)$, the charm quark contribution is partially negative, indicating destructive interference between the charm and light sectors. This interference arises from the relative orientation of the charm spin in the vector-heavy diquark versus the total pentaquark spin. Consequently, the light quark contribution exceeds 100\% of the total magnetic moment, while the charm quark reduces it. In $J_4(x)$, the charm quark contribution is negligible, and the light quarks entirely determine the magnetic moment.  
These results illustrate that the Dirac structure and diquark configurations of the interpolating currents directly influence which quark dominates the electromagnetic properties. The particularly striking result for $J_2(x)$—showing a large negative magnetic moment and unusual flavor decomposition—can be traced to its unique diquark configuration. Unlike the other currents, $J_2(x)$ contains a scalar-light diquark coupled to a vector-heavy diquark, which induces specific spin alignments that lead to strong destructive interference between the light and charm sectors. This configuration causes the individual quark contributions to partially cancel, resulting in $\mu_q$ and $\mu_c$ having opposite signs and explaining why $|\mu_q| > |\mu_{\text{tot}}|$. Currents containing scalar-heavy diquarks are optimal for probing charm quark magnetic moments, while vector-heavy diquark or antisymmetric constructions are more sensitive to light quark contributions. Moreover, the presence of destructive interference in $J_2(x)$ highlights the importance of considering both quark spin alignments and current structure when interpreting QCD sum rule predictions for pentaquark electromagnetic properties. Currents containing scalar-heavy diquarks are optimal for probing charm quark magnetic moments, while vector-heavy diquark or antisymmetric constructions are more sensitive to light quark contributions. Moreover, the presence of destructive interference in $J_2(x)$ highlights the importance of considering both quark spin alignments and current structure when interpreting QCD sum rule predictions for pentaquark electromagnetic properties. 
It should be emphasized that in some cases the flavor decomposition leads to ratios such as $\mu_q/\mu_{\text{tot}} > 100\%$ for the light-quark sector and $\mu_c/\mu_{\text{tot}} < 0$ for the charm-quark sector. These results arise from destructive interference between different quark contributions, rather than indicating any inconsistency or violation of physical principles. An important technical aspect of our flavor decomposition deserves clarification. The total magnetic moment is obtained as the sum of individual quark contributions, $\mu_{\text{tot}} = \mu_q + \mu_c$, due to the linearity of the electromagnetic current. The apparent cases where $\mu_q/\mu_{\text{tot}} > 100\%$ (as seen for $J_2(x)$) are physically permissible and indicate destructive interference between quark sectors—specifically, $\mu_q$ and $\mu_c$ have opposite signs, so their magnitudes can exceed that of the total moment. This phenomenon reflects the complex spin-charge correlations in multiquark systems and does not imply any violation of physical principles. In particular, the spin and charge factors associated with the light and charm quarks may combine in such a way that the charm contribution effectively reduces the total magnetic moment, while the light-quark part dominates. This feature is a natural consequence of multiquark dynamics in the QCD sum rule framework and provides additional information on the nontrivial internal spin–flavor structure of the pentaquark states.

\begin{widetext}

\begin{table}[htb!]
	\addtolength{\tabcolsep}{6pt}
	\caption{Flavor-decomposed magnetic moment contributions for the $P_{\psi}^{N}$ states and the relative ratios $\mu_{q}/\mu_{tot}$ and $\mu_{c}/\mu_{tot}$ (with sign, and in percentage). The results presented here have been extracted from the sum rules by fixing all input parameters at their central values.}
	\label{table3}
	\begin{ruledtabular}
\begin{tabular}{lccccccc}
	   \\
States & $\mu_{q} \,(\mu_N)$ & $\mu_{c} \,(\mu_N)$ & $\mu_{tot} \,(\mu_N)$ & $\mu_{q}/\mu_{tot}$ & $\mu_{c}/\mu_{tot}$ & $\mu_{q}/\mu_{tot} (\%)$ & $\mu_{c}/\mu_{tot} (\%)$ \\
	   \\
	   \hline\hline
$J_1(x)$ & $-0.02$  & $-1.08$  & $-1.10$ & $0.018$ & $0.982$   & $1.82$  & $~~98.18$ \\
$J_2(x)$ & $-5.61$ & $~~1.02$ & $-4.59$ & $1.222$ & $-0.222$ & $122.22$ & $-22.22$ \\
$J_3(x)$ & $-0.002$ & $-1.248$ & $-1.25$ & $0.002$ & $0.998$   & $0.16$  & $~~99.84$ \\
$J_4(x)$ & $-2.44$ & $~~0.00$ & $-2.44$ & $1.000$ & $0.000$  & $100.00$ & $~~0.00$ \\
\end{tabular}
\end{ruledtabular}
\end{table}

\end{widetext}

An important implication of our results is that the flavor-decomposed magnetic moments presented in Table~\ref{table3} provide a valuable benchmark for future ab initio studies, such as lattice QCD calculations. While QCD light-cone sum rules offer an analytical framework grounded in quark-hadron duality, lattice QCD can compute electromagnetic form factors directly from the QCD Lagrangian, providing an independent and rigorous check. Our analysis highlights several striking features that could be probed in such calculations: the large negative magnetic moment of the $J_2(x)$ configuration, the pronounced flavor hierarchy with the charm quark dominating the total moment in $J_1(x)$ and $J_3(x)$ while the light quarks dominate in $J_4(x)$, and the occurrence of destructive interference between quark sectors, particularly in $J_2(x)$. A lattice QCD investigation of these phenomena would require careful separation of connected and disconnected contributions from both light and heavy quarks. A convergence between lattice results and our predictions would provide strong support for the compact diquark structure assumed here, whereas any discrepancies would offer new insights into the interplay of quark configurations and electromagnetic properties in multiquark systems. These distinctive features therefore serve as clear, testable signatures, motivating further theoretical and computational studies of hidden-charm pentaquarks.

Finally, considering the mass values obtained using the $J_1(x)$ and $J_2(x)$ interpolating currents, it can be concluded that $J_1(x)$ predominantly couples to the $P_{\psi}^{N}(4312)$ state, whereas $J_2(x)$ is consistent with the $P_{\psi}^{N}(4457)$ state. Comparison with existing literature for states with the same quantum numbers provides a comprehensive perspective. In Ref.~\cite{Wang:2016dzu}, the magnetic moment of the $P_{\psi}^{N}(4312)$ state was calculated within the quark model, yielding $\mu_{P_{\psi}^{N}(4312)} = 1.76\,\mu_N$. Using QCD light-cone sum rules with a molecular configuration and $\rm{J^P = \frac{1}{2}^-}$, Ref.~\cite{Ozdem:2024jty} reported $\mu_{P_{\psi}^{N}(4312)} = 2.59^{+0.92}_{-0.81}\,\mu_N$, while Ref.~\cite{Li:2021ryu} predicted values in the range $1.624$–$1.737\,\mu_N$, accounting for coupled-channel effects and D-wave contributions. For the $P_{\psi}^{N}(4457)$ state, Ref.~\cite{Ozdem:2021ugy} obtained $\mu_{P_{\psi}^{N}(4457)} = 2.78^{+0.94}_{-0.83}\,\mu_N$ using a molecular pentaquark configuration within the QCD light-cone framework.  
The qualitative differences between our results and the quark-model prediction of Ref.~\cite{Wang:2016dzu} stem from fundamental differences in the theoretical treatment of the pentaquark's internal structure. In the quark model, the magnetic moment is essentially the sum of single-quark magnetic moments, $\mu_q \simeq e_q/(2 m_q^{\mathrm{eff}})$, weighted by a spatial wave function that is often symmetric or factorizable. This approach cannot capture the intricate spin‑charge correlations that emerge when quarks form compact, correlated diquarks with definite colour‑flavour‑spin symmetry—precisely the correlations built into our interpolating currents. 
Similarly, the discrepancy with molecular-model predictions (e.g., Refs.~\cite{Ozdem:2024jty,Li:2021ryu,Ozdem:2021ugy}) originates from the distinct spatial and spin configurations of the two pictures. Molecular models describe a weakly bound $\bar D^{(*)} \Sigma_c$ system, where the charm quark and antiquark reside in separate hadronic clusters. The magnetic moment therefore arises mainly from the sum of the constituent hadron moments, with only modest corrections from inter‑cluster exchange currents. In such a geometry, the charm‑quark spin is primarily coupled to the spin of the $\Sigma_c$ core, making strong anti‑alignment with the total pentaquark spin unlikely. In contrast, our compact diquark picture assumes all five quarks within a single confinement volume, with strong colour‑field correlations forming tightly bound diquark subunits. This leads to substantially different overlaps of the wave function with the electromagnetic current and to pronounced spin‑flavor correlations within the diquarks. 
The sign reversal observed for $J_1(x)$ and $J_2(x)$ relative to both quark‑model and molecular predictions is therefore a direct consequence of the compact diquark organization. In our QCD light‑cone sum‑rule analysis, the photon couples not only to individual quark charges but also to the collective spin of the diquark. For the axial‑vector heavy diquark $[uc]_{1^+}$ in $J_2(x)$, the charm‑quark spin can anti‑align with the total spin, producing a negative charm‑sector contribution—a correlation naturally encoded in the current structure but absent in both quark‑model wave functions and molecular hadron‑cluster descriptions. Moreover, non‑perturbative QCD effects (vacuum condensates, photon distribution amplitudes) further modify the electromagnetic vertices of the diquark subsystems. 
Consequently, the marked variation among the predictions of different approaches underscores that the magnetic moment is a sensitive discriminator of the internal quark‑gluon geometry and spin‑flavor organization. Our compact‑diquark scenario yields distinctive patterns—including negative moments and destructive inter‑flavor interference—that can be tested by future lattice QCD calculations and, ultimately, by experiment. 
Furthermore, the mass of the pentaquark state obtained with the current $J_3(x)$ lies slightly above the $\bar D \Lambda_c$ threshold, with values consistent within uncertainties. The corresponding magnetic moment can thus be compared with literature results, providing insight into potential agreements or discrepancies. In particular, Ref.~\cite{Ozdem:2023htj} predicted that $\mu_{\bar D \Lambda_c} = 0.44^{+0.17}_{-0.14}\,\mu_N,$ which differs significantly from the present result in both magnitude and sign. This discrepancy likely reflects differences in underlying quark configurations, choice of interpolating currents, and treatment of QCD dynamics. Overall, these findings indicate that the magnetic moments of hidden-charm pentaquark states are highly sensitive probes of their internal structures and spin–parity quantum numbers, emphasizing the importance of further theoretical and experimental investigations to achieve a more comprehensive understanding of their electromagnetic properties.

\section{Summary and conclusions}\label{summary}

In this work, we have systematically computed the magnetic moments of hidden-charm $P_{\psi}^{N}$ pentaquark states with $J^P = \frac{1}{2}^-$ using the QCD light-cone sum rule approach. Four distinct interpolating currents of the compact diquark–diquark–antiquark type were employed, revealing how the internal quark organization—particularly the scalar/axial-vector nature of diquark correlations—directly governs the electromagnetic response.  The methodology employed—including a careful analysis of OPE convergence, pole dominance, and systematic treatment of both perturbative and non-perturbative photon interactions—ensures robust numerical predictions. The significant variations observed across different currents highlight that magnetic moments are not merely static properties but dynamic probes of diquark spin alignment and internal quark dynamics.

Our principal findings are as follows:
\begin{itemize}
    \item The predicted magnetic moments span a considerable range: $\mu_{P_{\psi}^{N}} = -1.10^{+0.26}_{-0.22}\,\mu_N$ for $J_1(x)$, $-4.59^{+0.98}_{-0.75}\,\mu_N$ for $J_2(x)$, $-1.25^{+0.30}_{-0.26}\,\mu_N$ for $J_3(x)$, and $-2.44^{+0.62}_{-0.45}\,\mu_N$ for $J_4(x)$. This variation directly reflects the distinct diquark configurations and mass spectra of the corresponding states, with $J_1(x)$ and $J_2(x)$ coupling predominantly to $P_{\psi}^{N}(4312)$ and $P_{\psi}^{N}(4457)$ respectively, while $J_3(x)$ and $J_4(x)$ represent alternative structural arrangements with masses near 4.20–4.25 GeV.
    
    \item Flavor decomposition uncovers strikingly different quark-sector contributions: charm quarks dominate ($\sim$98–99\%) in the scalar-diquark dominated currents $J_1(x)$ and $J_3(x)$, while light quarks prevail in $J_4(x)$ (100\%). Most notably, the $J_2(x)$ current exhibits destructive interference, with light quarks contributing 122\% and charm quarks –22\% of the total moment. This pattern highlights the crucial role of spin alignment in vector diquark systems, where specific charm quark orientations within the vector-heavy diquark lead to partial cancellation between quark sectors.
    
    \item These results differ markedly from molecular-model predictions in both magnitude and sign. For instance, while molecular approaches typically yield positive moments around $+2.6$–$+2.8\,\mu_N$ for $P_{\psi}^{N}(4312)$ and $P_{\psi}^{N}(4457)$, our compact diquark picture predicts negative values, suggesting that compact and molecular pictures produce fundamentally different electromagnetic responses due to their distinct spatial wavefunctions and spin-correlation patterns.
\end{itemize}

Looking forward, our results provide concrete benchmarks for future lattice QCD simulations and underscore the importance of electromagnetic observables in unraveling the nature of exotic hadrons. Experimental progress in measuring these quantities, though challenging, would offer direct insight into the quark–gluon architecture of pentaquark states. We anticipate that continued theoretical and experimental efforts along these lines will be crucial for achieving a comprehensive understanding of the $P_{\psi}^{N}$ states and their role in the spectrum of exotic hadrons.


 \newpage
 
  \appendix
  \begin{widetext}
  
  \section*{APPENDIX: EXPLICIT FORMS OF THE SUM RULES FOR $\rho_i (\mathrm{M^2},\mathrm{s_0})$}
This appendix presents the explicit analytical expressions derived for the magnetic dipole moments of the $P_{\psi}^{N}$ pentaquarks. The related expressions are written as follows:
   \begin{align}
\label{F1sonuc}
 \rho_1 (\mathrm{M^2},\mathrm{s_0})&=-\frac {61\ e_c } {2^{25} \times 3  \times 5^3 \times 7^2 \pi^7}  I[0,7]\nonumber\\
        &+\frac { e_u m_c \langle g_s^2G^2\rangle  \langle \bar q q \rangle} {2^{24} \times 3^6 \times 5 \pi^5} \Big[ (40 \mathbb A[u_0] + 207 I_3[\mathcal S ]) I[0, 3] - 16 \chi I[0, 4] \varphi_\gamma[u_0]  \Big]\nonumber\\
       &+\frac { f_{3\gamma}\langle g_s^2G^2\rangle } {2^{24} \times 3^6 \times 5 \pi^5}  \Big[ (e_u-9 e_d ) I[0, 4] \psi^a[u_0]\Big]\nonumber
       \nonumber\\
        &+\frac {  e_u m_c\langle \bar q q \rangle} {2^{25} \times 3 \times 5^2 \times 7 \pi^5}  I_3[\mathcal S] I[0, 5], \\
 \rho_2 (\mathrm{M^2},\mathrm{s_0})&=\frac {61} {2^{25}\times 3 \times 5^2 \times 7^2 \pi^7} 
  (2 e_c - 9 e_u )I[0,7]\nonumber\\
        &+\frac {e_u\, \langle g_s^2G^2\rangle  \langle \bar q q \rangle} {2^{23} \times 3^5 \times 5 \pi^5} \Big[ (40 \mathbb A[u_0] + 75 I_3[\mathcal S] - 216 I_3[\mathcal{\tilde S}]) I[0, 3] - 
  52 \chi I[0, 4] \varphi_\gamma[u_0] \Big]\nonumber\\
       &-\frac { f_{3\gamma}\langle g_s^2G^2\rangle } {2^{28} \times 3^6 \times 5 \pi^5}  \Big[ 
         (1035 \,e_u   I_1[\mathcal V] + 16 (9 e_d + 55 e_u) \psi^a[u_0])I[0, 4]\Big]
       \nonumber\\
               &+\frac {  m_c  \langle \bar q q \rangle} {2^{22} \times 3 \times 5^2 \times 7 \pi^5} (e_d+e_u)\Big[ (7 \mathbb A[u_0] + 20 I_3[\mathcal S] - 24 I_3[\mathcal{\tilde S}]) I[0, 5] -  2 \chi  \varphi_\gamma[u_0] I[0, 6]\Big]\nonumber\\
               &+\frac {e_u\, f_{3\gamma} } {2^{24} \times 3^2 \times 5^2 \times 7 \pi^5}  \Big[
               (-35 I_1[\mathcal V] + 258 \psi^a[u_0])I[0, 6] \Big].  \label{F2sonuc}\\
        \rho_3 (\mathrm{M^2},\mathrm{s_0})&=-\frac {61\ e_c } {2^{24} \times 5^3 \times 7^2 \pi^7}  I[0,7]\nonumber\\
        &+\frac { m_c \langle g_s^2G^2\rangle  \langle \bar q q \rangle}  {2^{22} \times 3^6 \times 5 \pi^5} (2 e_d+e_u)\Big[   \mathbb A[u_ 0] I[0, 3] - 45 I_ 3[\mathcal {\tilde S}] I[0, 3] + 
 4 \chi I[0, 4] \varphi_\gamma[u_ 0]  \Big]\nonumber\\
       &-\frac { f_{3\gamma}\langle g_s^2G^2\rangle } {2^{24} \times 3^6 \times 5 \pi^5}  (2 e_d+e_u) \Big[  I[0, 4] \psi^a[u_0]  \Big]\nonumber
       \nonumber\\
        &+\frac { m_c\langle \bar q q \rangle} {2^{25} \times 3 \times 5^2  \pi^5}  (2 e_d+e_u) I_3[\mathcal S] I[0, 5], \label{F3sonuc}\\
        \rho_4 (\mathrm{M^2},\mathrm{s_0})&=-\frac {183} {2^{26}\times 5^3 \times 7^2 \pi^7} 
   (e_d + 5 e_u) I[0,7]\nonumber\\
        &+\frac { m_c \langle g_s^2G^2\rangle  \langle \bar q q \rangle} {2^{22} \times 3^4 \times 5 \pi^5} (e_d + 5 e_u)\Big[ (\mathbb A[u_0] + 13 I_3[\mathcal S]) I[0, 3] - \chi \varphi_\gamma[u_0]  I[0, 4]  \Big]\nonumber\\
        &-\frac { f_{3\gamma}\langle g_s^2G^2\rangle } {2^{27} \times 3^4 \times 5 \pi^5} (e_d + 5 e_u)  \Big[   (23  I_1[\mathcal V] + 24  \psi^a[u_0])I[0, 4] \Big]
       \nonumber\\
        &+\frac {   m_c\langle \bar q q \rangle} {2^{22} \times 3 \times 5  \pi^5} (e_d + 5 e_u) I_3[\mathcal S] I[0, 5]\nonumber\\
       &-\frac { f_{3\gamma} } {2^{25} \times 3^2 \times 5^2 \times 7 \pi^5} (e_d + 5 e_u)  \Big[   (35 I_1[\mathcal V] - 258 \psi^a[u_0]) I[0, 6] \Big].  \label{F4sonuc}
              \end{align} 
Here, $\varphi_\gamma(u)$, $\psi^v(u)$, $\psi^a(u)$, ${\cal A}(\alpha_i)$, ${\cal V}(\alpha_i)$, $h_\gamma(u)$, $\mathbb{A}(u)$, ${\cal S}(\alpha_i)$, ${\cal \tilde S}(\alpha_i)$, ${\cal T}_1(\alpha_i)$, ${\cal T}_2(\alpha_i)$, ${\cal T}_3(\alpha_i)$, and ${\cal T}_4(\alpha_i)$ denote the photon DAs. Their explicit expressions, together with the required numerical parameters, are presented in Ref.~\cite{Ball:2002ps}. The functions $I[n,m]$ and $I_i[\mathcal{F}]$ are defined as follows:
\begin{align}
 I[n,m]&= \int_{4m_c^2}^{\mathrm{s_0}} ds ~ e^{-s/\mathrm{M^2}}~
 s^n\,(s-4m_c^2)^m,\nonumber\\
 I_1[\mathcal{F}]&=\int D_{\alpha_i} \int_0^1 dv~ \mathcal{F}(\alpha_{\bar q},\alpha_q,\alpha_g)
 \delta'(\alpha_ q +\bar v \alpha_g-u_0),\nonumber
   \end{align}
 \begin{align}
  I_2[\mathcal{F}]&=\int D_{\alpha_i} \int_0^1 dv~ \mathcal{F}(\alpha_{\bar q},\alpha_q,\alpha_g) \delta'(\alpha_{\bar q}+ v \alpha_g-u_0),\nonumber\\
     I_3[\mathcal{F}]&=\int D_{\alpha_i} \int_0^1 dv~ \mathcal{F}(\alpha_{\bar q},\alpha_q,\alpha_g)
 \delta(\alpha_ q +\bar v \alpha_g-u_0),\nonumber\\
   I_4[\mathcal{F}]&=\int D_{\alpha_i} \int_0^1 dv~ \mathcal{F}(\alpha_{\bar q},\alpha_q,\alpha_g) \delta(\alpha_{\bar q}+ v \alpha_g-u_0).
 \end{align}
 Here, $ \mathcal{F} $ represents the corresponding photon DAs.

It is worth emphasizing that the Borel transformations applied in the above expressions are performed according to the following relations:
\begin{align}
 \mathcal{B}\bigg\{ \frac{1}{\big[(p^2-m_i^2)\big]\big[(p+q)^2-m_f^2\big]} \bigg\} &\;\longrightarrow\; e^{-m_i^2/\mathrm{M_1^2} - m_f^2/\mathrm{M_2^2}},
\end{align}
on the hadronic side, and
\begin{align}
 \mathcal{B}\bigg\{ \frac{1}{\big(m^2 - \bar u p^2 - u(p+q)^2 \big)^{\alpha}} \bigg\} &\;\longrightarrow\; (\mathrm{M^2})^{(2-\alpha)} \, \delta(u-u_0)\, e^{-m^2/\mathrm{M^2}},
\end{align}
on the QCD side, where
\begin{align*}
 \mathrm{M^2} = \frac{\mathrm{M_1^2} \mathrm{M_2^2}}{\mathrm{M_1^2}+\mathrm{M_2^2}}, 
 \qquad
 u_0 = \frac{\mathrm{M_1^2}}{\mathrm{M_1^2}+\mathrm{M_2^2}}.
\end{align*}
Here, $\mathrm{M_1^2}$ and $\mathrm{M_2^2}$ denote the Borel parameters associated with the initial and final $P_{\psi}^{N}$ states, respectively. Since the same $P_{\psi}^{N}$ states appears in both channels, we adopt the choice $\mathrm{M_1^2} = \mathrm{M_2^2} = 2\mathrm{M^2}$ and $u_0 = \tfrac{1}{2}$. This prescription ensures that the single-dispersion approximation efficiently suppresses contributions from higher resonances and continuum states. Further technical details of this procedure can be found in Ref.~\cite{Ozdem:2024dbq}.

  \end{widetext}

\bibliographystyle{elsarticle-num}
\bibliography{PccbarMDM.bib}

\end{document}